\documentclass[longbibliography,%
 reprint,
superscriptaddress,
 amsmath,amssymb,
 aps,
prl,
]{revtex4-1}

\usepackage{graphicx}
\usepackage{dcolumn}
\usepackage{bm}
\usepackage{color}
\usepackage{soul}
\usepackage{hyperref}

\definecolor{darkgreen}{rgb}{0.0, 0.5, 0.0}

\newcommand{\avg}[1]{\langle #1\rangle}

\newcommand{\fref}[1]{\textcolor{black}{Fig.~\ref{fig:#1}}}

\newcommand{\eref}[1]{Eq.~\ref{eqn:#1}}

\newcommand{\elabel}[1]{\label{eqn:#1}}

\begin{document}

\preprint{APS/123-QED}

\title{Exploiting bias in optimal finite-time copying protocols}

\author{Daan Mulder}
 \email{d.mulder@amolf.nl}
\affiliation{AMOLF, Science Park 104, 1098 XG, Amsterdam, The Netherlands}

\author{Pieter Rein ten Wolde}
\affiliation{AMOLF, Science Park 104, 1098 XG, Amsterdam, The Netherlands}

\author{Thomas E. Ouldridge}
\affiliation{Department of Bioengineering, Imperial College London, London SW7 2AZ, United Kingdom}

\date{\today}

\begin{abstract}
  We study a finite-time cyclic copy protocol that creates
  persisting correlations between a memory and a data bit. The average
  work to copy the two states of the data bit consists of the
  mutual information created between the memory and data bit after
  copying, a cost due to the difference between the initial and final
  states of the memory bit, and a finite-time cost. At low copy
  speeds, the optimal initial distribution of the memory bit matches
  the bias in the expected outcome, set by the bias in the data bit
  and the copying accuracies. However, if both states of the data are
  copied with the same accuracy, then in the high-speed regime copying
  the unlikely data bit state becomes prohibitively costly with a
  biased memory; the optimal initial distribution is then pushed
  towards 50:50. Copying with unequal accuracies, at fixed
  copy-generated mutual information, yields an opposite yet more
  effective strategy. Here, the initial memory distribution becomes
  increasingly biased as the copy speed increases, drastically
  lowering the work and raising the maximum speed. This strategy is so effective that it induces a symmetry breaking transition for an unbiased data bit.
\end{abstract}

\maketitle

\noindent The energetic cost of computing is a growing concern, as
energy consumption will become the main factor limiting the growth of
worldwide computing capacity in the coming decade
\cite{semiconductors21}.  At heart, computing concerns manipulating
information-bearing degrees of freedom, which has a fundamental
thermodynamic cost \cite{landauer61, szilard29}.  This cost is
typically studied in the context of a bit reset, which in the reversible limit
 requires
$k_{\rm B} T \ln(2)$ of work to reduce the entropy of the
bit~\cite{berut.2012,Koski.2014tu,Jun.2014qmn}.  If current trends in
transistor efficiency continue, we will approach this Landauer bound
halfway the 21st century~\cite{wong17}.

A ubiquitous motif in both man-made and natural systems is copying the
state of a data bit into a memory bit \cite{Pigolotti13,Parrondo15,
  peliti2021, bennett82,bennett88, ueda09, tenwolde17,tenwolde19}. During the copy step, the
state of the memory bit becomes correlated with the state of the data
bit via an interaction energy, but after this step the memory bit is physically uncoupled from the
data bit yet remains correlated. The
combined system is therefore out of thermodynamic equilibrium, such
that copying  requires free energy input.

\begin{figure}
    \centering
    \includegraphics[width=\columnwidth]{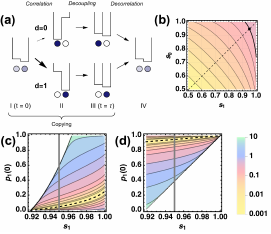}
    \caption{Copying a bit. (a) Copy protocol. Double-well potential showing energy $E$ for memory state $m=0$ (left) and $m=1$ (right). Circles below show respective occupancies. A black line between the wells implies memory cannot switch. The upper row corresponds to copying  data bit state $d=0$; lower row to $d=1$. (b)  Contour plot of $I(D;M)$ in $s_0$-$s_1$-plane for $P(D=1) =0.9$. Contour lines are asymmetric due to bias data bit. The bold contour crosses dashed equal-accuracy line at $s_0=s_1 = 0.95$. Contour plots of $W_0$ (c) and $W_1$ (d) as function of initial memory distribution $p_1(0)$ and copy accuracy $s_1$, for $\tau=2.5$, with $s_1$ and $s_0$ covaried along bold iso-information line in (b). Black dashed lines give $p_1(0)=p_1(\tau)$ for which $W_d = 0$, with $p_1(\tau) = s_1$ for $d=1$ $p_1(\tau) = 1-s_0$ for $d=0$.  Grey lines mark $s_0=s_1=0.95$.}
    \label{fig:measurementfinitetime}
\end{figure}

To understand the thermodynamic cost of copying, we must consider
  a full {\em cycle} \cite{bennett82,bennett88, ueda09, tenwolde17,
    tenwolde19} (see Fig. \ref{fig:measurementfinitetime}). Let us
  consider an unbiased data bit, and a memory whose states are equally
  stable. In Bennett's pioneering work \cite{bennett82}, the cycle of
  copying this data bit starts with the memory in a well defined state
  0, and the free energy stored in this non-equilibrium state is used
  to pay for the copy operation; indeed, this step can be performed
  without external work input. Yet, to complete the cycle, the memory
  bit must be reset to its original non-equilibrium state and this
  ``reset'' or ``erasure'' step requires external work
  input. Importantly, this choice is neither fundamental nor necessary
  \cite{ueda09}. The cycle can also start with the memory in a state
  that is unbiased, either because the outcome of the previous copy
  operation is overwritten directly, akin to cellular copying
  \cite{tenwolde17,tenwolde19}, or because the memory has been allowed
  to freely equilibrate before the copy step. In either case, the
  correlation-generating copy step requires external work input but
  there is no work consuming reset. Crucially, while these protocols
  differ in the steps that require work input, the memory manipulation
  involves the same work input over the full cycle in the quasi-static
  limit. For typical copy protocols, in which the data bit does not evolve during the measurement,
  this net work input implies thermodynamic irreversibility.

A full copy cycle is, however, not necessarily
irreversible. The free energy stored in the correlations could be
harvested by reversing the correlation step in a reversible manner, by
slowly bringing the memory back into contact with the same data
bit.   
Indeed, the fundamental reason why copy cycles tend to be irreversible, even when performed quasi-statically,
is the failure to recover the free energy stored in the correlations
after the copy step  \cite{ueda09,tenwolde17}.
This also holds true for
copy cycles that include a reset step. Resetting (or erasing) a memory bit that is correlated with the data bit will destroy correlations. If the reset protocol is data agnostic, it will be thermodynamically irreversible; reversing this protocol cannot recover the correlated state, and the stored free energy is lost. If, however,  resetting is performed after carefully decorrelating the bits, it can be thermodynamically reversible. 
One can first reverse the correlation step by bringing the memory into and out of contact with the data, 
and then reset. Reversing this procedure in turn would recover the correlated state and no free energy will be lost. 
It is therefore the failure to extract the free energy stored in the correlations \cite{ueda09,tenwolde17}, rather than erasing or resetting per se,
that explains the minimal cost of copy cycles.

Harvesting the free energy stored in the correlations between memory and data bit by reusing the data bit is typically not feasible. 
This  free energy then provides a lower bound on the cost of a full copy cycle, which
  is only reached if the operations are performed in a quasi-static,
  thermodynamically reversible manner. Yet, copy operations typically
  need to be performed in finite time, such that the system must be
  driven out of thermal equilibrium, raising the work beyond the
  quasi-static bound. While bit reset in finite time
\cite{Esposito2013, peliti2013, Jun.2014qmn,deweese14,bechhoefer20,dahlsten21,
  scandi22, park22,Vu.2023} and a copy cycle in the quasi-static limit
\cite{bennett82,bennett88,ueda09, tenwolde17,tenwolde19} have been
well studied, a full copy cycle in finite time has received little
attention \cite{Nagase2024}.  Many questions therefore remain
unanswered.  In particular, we anticipate a trade-off between
minimising the work to copy the two respective states of the data bit
individually.  What is therefore the optimal protocol that minimizes
the average work to copy the data bit?  More specifically, what is the optimal initial distribution of the memory bit? In the absence of a time limit on the operation, whether the memory is reset to 0 at the start of the copy step as in \cite{bennett82, Nagase2024}, is unbiased, or something intermediate is irrelevant for calculating the lower bound on the total cost of the cycle; given infinite time, these initial distributions can all be interconverted at no cost. We anticipate that this equivalence no longer holds when the copy operation must occur in finite time. How does this trade-off depend
on the bias in the data bit and the copying speed? If the copy
operation needs to generate a desired mutual information between the
memory and data bit, can the relative accuracy of copying the two data
bit states be leveraged to lower the work? How does the optimal
protocol depend on the steps that are under a time constraint
(Fig. \ref{fig:measurementfinitetime})?  In this manuscript, we
address these questions, modelling the memory bit as a two-state
system with constrained switching rates.

The average work per copy can be written as a sum of three terms: the created mutual information between the memory and data bits, the free energy stored in the memory bit alone, and the finite-time cost. The second term, decisive at low copy speeds, favors an initial distribution of the memory that matches the distribution of the copy outcome, set by the bias in the data bit and the copying accuracies. This distribution also minimizes the finite-time costs in the low-speed regime. However, this matching strategy would give prohibitively large finite-time costs in the high-speed regime, if both data bit states are copied with the same accuracy.
Instead, the optimal initial memory distribution is pushed towards 50:50, even when the data bit is biased. A more effective strategy is to increase the accuracy of copying the likely state of the data bit and lower that of the unlikely state. This approach drastically lowers the work and raises the maximum copy speed. Moreover, it yields more extreme biases in the optimal initial memory distribution than present in the data, in sharp contrast to the equal accuracy scenario. Remarkably, when the data bit is unbiased, this strategy can even generate a symmetry breaking transition, where the optimal initial memory distribution is pushed to either 0 or 1. 
  
\textit{Copy cycle - } We model our memory M as a two-state device, with transition rates  $k_{ij}(t)$ to go from state $j$ to $i$. The master equation of its probability distribution $P_M = \{p_0,p_1\}$ is
\begin{align}
    \dot{p_1} = - \dot{p_0} = k_{10}(t)p_0(t) - k_{01}(t)p_1(t).
\label{eq:mastereq}
\end{align}
The energy levels and rate constants obey $\Delta E(t) \equiv E_1(t)-E_0(t) =  \beta^{-1} \ln(k_{01}/ k_{10})$, so that the equilibrium state obeys the Boltzmann distribution.  When the states are separated by a large barrier the rates $k_{ij}$ are negligibly small. During the copy process the barrier is lowered and transitions can occur as the energy levels are manipulated to drive the copying. To study copying in finite time, however, a constraint on the transition rates is required; we assume a constant relaxation rate $k_T = k_{10}+k_{01}$ \cite{Esposito_2010, Esposito2013}, independent of $\Delta E$. In addition, $\beta^{-1} = k_B T = 1$ and $k_T=1$. We will study a cyclic protocol and set $E_0 = 0$; the power is thus $\dot{W} = p_1 \dot{E_1} $.

The full cycle consists of three subprocesses  (\fref{measurementfinitetime}(a)). As we motivate below, we start the cycle with the memory in an equilibrium state, with the barrier between the two states lowered (configuration I). The initial energy $E_1(t=0)$ thus sets the initial distribution $P^i_M$, parametrised by $p_1(0)$, which is a key degree of freedom that will be optimized to lower the work. During the \textit{correlation step} of duration $\tau$ the memory bit is brought in contact with the data bit, which changes its energy and hence its distribution from $p_1(0)$ (configuration I) to $p_1(\tau)$, which depends on the state of the data bit $d$ (configuration II). Specifically, 
  after this correlation step
$p_1(\tau) = 1-s_0$ when the state of the data bit $d=0$ and $p_1(\tau) = s_1$ when $d=1$,
  where $s_0$ and $s_1$ are the accuracies of copying the respective states of the data bit; $s_0,s_1=1$ implies a perfect copy, whereas $s_0,s_1=1/2$ implies no correlations are generated. We consider data bits with bias $P(D=1)\equiv p'$, and, to meaningfully compare protocols, copying at fixed final mutual information $I(D;M)$ between memory and data \cite{SI}. As \fref{measurementfinitetime}(b) shows, a given $I(D;M)$ corresponds, for a given $p'$, to a range of possible values of $s_0$ and $s_1$. 

The copy must be able to persist regardless any subsequent changes in
the data.  Hence, the memory and data must be decoupled, while the
memory state is retained.  We model this property by adding a
\textit{decoupling step}. In this step the energy barrier between
  the two states is raised, which fixes the probability distribution
  of the memory $p_1(\tau)$; the memory is decoupled from the data bit, which resets its energy to $E_1(\tau)=E_1(0)$ independent of the data bit state
  (configuration III). It is this configuration of the memory
  that stores the state of the data, since the two are both correlated
  and decoupled.

To complete the cycle and make the memory ready for the next copy cycle, we bring the memory distribution back to its initial state. In this {\em decorrelation step} the correlations created in the copy process are destroyed.

We now address how at the start of the cycle we choose the memory's resting energy $E_1(\tau)=E_1(0)$ and how in the decorrelation step we bring the memory distribution back to its original state (\fref{measurementfinitetime}(a)).   There are two key observations to answer these questions: a)  the {\em marginal} distribution of the memory after the copy operation, averaged over the two states of the data $d=0,1$, is fixed; it is fully specified by the statistics of the data and the copy accuracies: $\avg{p_1(\tau)} = (1-p')(1-s_0)+p' s_1$; b)  as we show below, there exists an optimal distribution of the memory at the beginning of the copy operation, $p_1^{\rm opt}(0)$, that minimizes the overall work, which, in general, differs from $\avg{p_1(\tau)}$. The question then becomes how to change the memory distribution during the decorrelation step from $\avg{p_1(\tau)}$ to $p_1^{\rm opt}(0)$.
  This  reinitialization could be achieved by  bringing the memory back into contact with the data bit, which would in principle make it possible to recover all the stored free energy, yielding in the quasi-static limit a fully reversible cycle; however, this data-dependent decorrelation step is typically unfeasible since it would require the data to remain unchanged.
  
 One procedure that is independent of the data bit and that guarantees that $\avg{p_1(\tau)}$ is moved back to the optimal $p_1^{\rm opt}(0)$ is to set $E_1$ in the decoupling step to the value for which the desired $p_1^{\rm opt}(0)$ is the equilibrium distribution, $E_1(\tau)=E_1(0) = \ln((1-p_1^{\rm opt}(0))/p_1^{\rm opt}(0))$, and then proceed in the decorrelation step by simply lowering the barrier, and giving the distribution ample time to relax from $\avg{p_1(\tau)}$ to $p_1^{\rm opt}(0)$; this is the procedure shown in \fref{measurementfinitetime}(a), where the memory begins and ends in an equilibrium state.  While this procedure I  is the simplest, it does not recover any of the work that was performed during the copy operation.
 We also consider two other procedures, which differ in the decorrelation step. In the decorrelation step of procedure II,
 $E_1$ is instantly set to a level $\ln((1-\avg{p_1(\tau)}))/\avg{p_1(\tau)})$ for which $\avg{p_1(\tau)}$ is the equilibrium distribution, the barrier is lowered, and then $E_1$ is quasi-statically changed back to the value $E_1(0)$ for which the desired $p_1^{\rm opt}(0)$ is the equilibrium distribution. As we will see, this procedure has the smallest overall cost because it will recover  the work that is stored in the marginal distribution of the memory bit after the copy operation. In procedure III,  the decorrelation step is omitted altogether and, in the next cycle, the memory is directly overwritten by the new data bit to be copied. In this case, the initial distribution of the memory is set by its marginal distribution after the last copy operation, which is, as we will see, in general suboptimal.

It is possible to add operations between the
  decorrelation step and the start of the next copy cycle,
  e.g. bringing the memory to one of the two states or setting it
    to an even, 50:50, distribution. However, these embellishments
  can never decrease the work and have therefore been omitted. We
  will quantify procedure I in the main text, and present the
  details of the other two in the SI.

We start by deriving the minimally required average work to generate a desired mutual information $I(D;M)$ between the memory and data bit during the copy operation of time $\tau$  (from I to III in Fig. \ref{fig:measurementfinitetime}). 
This minimization requires two distinct optimization procedures. 
First, we derive the minimal work $W_d$ to copy each state $d=0,1$ of the data bit separately, given an initial distribution of the memory $p_1(0)$ and accuracy $s_1$ (which sets $s_0$ since we fix $I(D;M)$, see \fref{measurementfinitetime}(b)). 
Next, using the distribution of the data bit (parametrized through $p'$) we find the optimal $p_1(0)$ and $s_1$ that minimize the average cost per copy $\mathcal{W} =  (1-p') W_0 + p' W_1$.

\textit{Copying a data bit state -}
To obtain the minimal work $W_d$ to copy the data bit in state $d$ in time $\tau$, we integrate the power $\dot{W} = p_1 \dot{E_1}$ over one row in Fig. \ref{fig:measurementfinitetime}(a), yielding
\begin{align}
W_d = \Delta F_d + T \Delta S_d^\text{irr},
\label{eq:deltafandsirr}
\end{align}
where $\Delta F_d = \Delta U_d - T \Delta S_d$ is the Landauer-like cost, and $\Delta S_d^\text{irr}$ is the minimal finite-time cost  (see \cite{SI}). We minimize the work by minimizing the finite-time cost via a Lagrangian \cite{Esposito_2010, Esposito2013}. This calculation is akin to a bit reset, and yields the optimal protocol for bringing the memory from its initial to final distribution.

\begin{figure}
    \centering
    \includegraphics[width=\columnwidth,trim={0 0.5cm 0 0}]{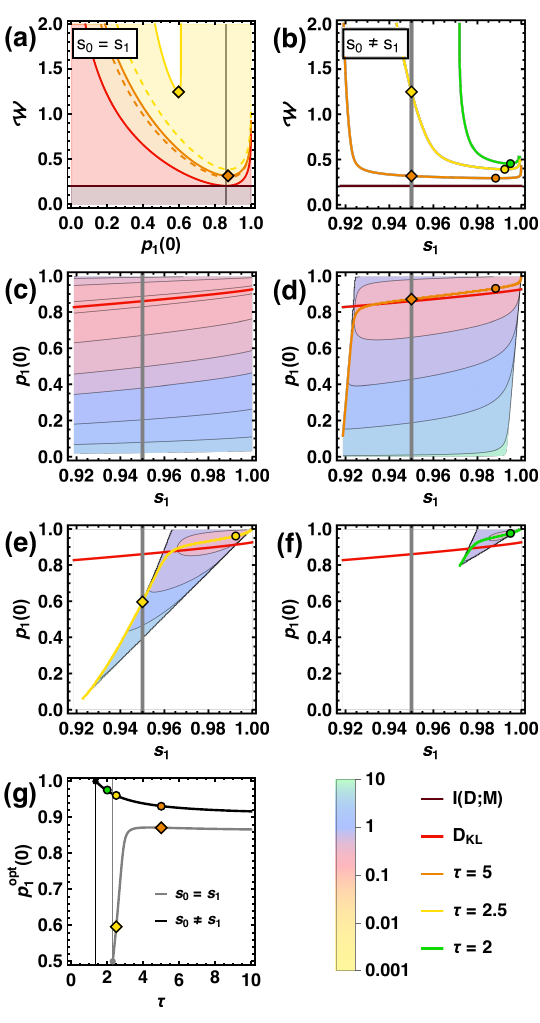}
    \caption{Average work along the bold iso-information contour in Fig.~\ref{fig:measurementfinitetime}(b), with $p'=0.9$. (a) Work $\mathcal{W}$ (\eref{Wtot}) as a function of $p_1(0)$ for $s_0=s_1=0.95$, and $\tau=2.5$ (yellow),$5$ (orange). Contributions shown as areas under the curve.  Dashed vertical line denotes $p_1(0) = \avg{p_1(\tau)}$, where   $D_{KL}(P^f_M || P^i_M)=0$. Diamonds mark $p_1^{\rm opt}(0)$ that minimizes $\mathcal{W}$, which moves towards $1/2$ for small $\tau$. Dashed lines denote $1/\tau$ approximation. (b) $\mathcal{W}$ as function of $s_1$, for different $\tau$; $p_1(0)$ has been optimized for each $s_1$, and corresponding $p_1^{\rm opt}(0)$ is shown in (d)-(f) by lines in corresponding color. Circles mark overall minimum of $\mathcal{W}$; diamonds mark minimum for $s_0 = s_1$. Unequal copying accuracies lower the work and raise maximum copy speed. 
      (c)  Quasi-static work $I(D;M) + D_{KL}(P^f_M || P^i_M)$, minimized when $p_1(0) = \avg{p_1(\tau)}$ (red line). Work $\mathcal{W}$ for $\tau=5$ (d), $\tau=2.5$ (e) and $\tau =2$ (f). Unequal copying accuracies allow for copying at smaller $\tau$. Grey vertical lines mark $s_0=s_1=0.95$. (g) $p_1^{\rm opt}(0)$ as function of $\tau$. For large $\tau$, $p_1^{\rm opt}(0)$ reflects bias data for both equal and unequal copying accuracies; for short $\tau$, $p_1^{\rm opt}(0)\to 1/2$ for equal yet $p_1^{\rm opt}(0) \to 1$ for unequal accuracies. Vertical lines mark $\tau_{\rm min}$. }
    \label{fig:main}
\end{figure}

Fig. \ref{fig:measurementfinitetime}(c,d) shows that, for a given generated mutual information $I(D;M)$, the minimal work $W_d$ to copy the two respective states of the data bit $d=0,1$ depends on the initial distribution of the memory and the relative copying accuracy. The work $W_d$ to copy a given state $d$ is zero when the initial memory distribution $p_1(0)$ equals the final one $p_1(\tau)$, which is determined by the data bit state $d$ and the copying accuracies: $p_1(\tau) = s_1$ if $d=1$ and $p_1(\tau) = 1-s_0$ if $d=0$, marked by the black dashed lines.  At low speeds, the work approaches the quasi-static cost $\Delta F_{d}$ as $1/\tau$ \cite{SI}, in agreement with earlier work on resetting \cite{deweese14, Jun.2014qmn, aurell11, ito21}. The cost rises non-linearly with the distance between $p_1(0)$ and $p_1(\tau)$, and this rise is faster for higher copy speeds \cite{SI}. The cost diverges for 
$\tau_\text{min}= \ln(\text{max}(p_1(0)/p_1(\tau), (1-p_1(0))/(1-p_1(\tau))))$,
 which is the origin of the inaccesible white regions. In these regions, the distance over which the memory distribution must be moved is too large for the memory's finite transition rates. A $100\%$ accuracy can not be reached within finite time, and every accuracy comes with a minimal required time \cite{Esposito2013, dahlsten21}. Last but not least, changing $p_1(0)$  to minimize $W_1$ tends to increase $W_0$, and {\it vice versa}. This trade-off suggests there exists an optimal $p_1(0)$  that minimizes the average cost ${\cal W}$.

\textit{Average work -} Using Eq. (\ref{eq:deltafandsirr}) for $W_d$, the average work can be decomposed into three terms \cite{ueda09, Nagase2024, SI}:
\begin{align}
   \mathcal{W} =  k_B T I(D;M) + k_B T D_{KL}(P^f_M || P^i_M) + T \Delta \mathcal{S}^\text{irr}.
    \elabel{Wtot}
\end{align}
Here, $I(D;M)$ is the mutual information between data and memory after copying. 
It depends on the data bias $p'$ and the accuracies $s_0,s_1$. Since we consider copying at a given $I(D;M)$ this term cannot be optimized. 
The second term, with $D_{KL}(P^f_M || P^i_M)$ the Kullback-Leibler divergence between the memory distributions before and after the copy step, equals the non-equilibrium free energy stored in the memory bit alone by the copy process. The final distribution $P^f_M$, parametrised by $\avg{p_1(\tau)}$, is fixed by the data bias and copying accuracies: $\avg{p_1(\tau)}=(1-p')(1-s_0)+p's_1$. However, the initial memory distribution  $P^i_M$, set by $p_1(0)$, can be optimized:  when it is chosen to match the final one such that $p_1(0) = \avg{p_1(\tau)}$, $D_{KL}(P^f_M || P^i_M)=0$, its minimal value.

Yet, to minimize the total work $\cal{W}$, we also need to address the third term in \eref{Wtot}. 
The term $\Delta \mathcal{S}^\text{irr}$ is the minimal finite-time cost, averaged over both states of the data bit. 
Unlike $I(D;M)$ and $D_{KL}(P^f_M || P^i_M)$, it depends on the copy time $\tau$, as well as $p_1(0)$ and $s_0,s_1$. For large values of $\tau$, the minimum of $\Delta \mathcal{S}^\text{irr}$ is close to $\avg{p_1(\tau)}$, the minimum of the $D_{KL}$-term. However, this is no longer the case in the short-time regime. The result is a trade-off between  $D_{KL}(P^f_M || P^i_M)$ and $\Delta \mathcal{S}^\text{irr}$, with the optimal balance between them set by the copy time $\tau$.

To illustrate the trade-off between  $D_{KL}(P^f_M || P^i_M)$ and $\Delta \mathcal{S}^\text{irr}$, we first consider the special case where the two data bit states are copied with the same accuracy: $s_0=s_1=s$. 
Figure \ref{fig:main}(a) shows $\mathcal{W}$ for two different values of the copy time $\tau$ as a function of $p_1 (0)$, for $s = 0.95$.
The mutual information $I(D;M)$ is indeed independent of $p_1(0)$ and $\tau$. 
In contrast, $D_{KL}(P^f_M || P^i_M)$ is minimized when the initial distribution equals the final one: $p_1(0) = \avg{p_1(\tau)}$ (vertical line). 
This value of $p_1(0)$ also minimizes $\mathcal{W}$ in the regime where $\tau$ is large, not only because  $\Delta \mathcal{S}^\text{irr}$ is relatively small in this regime, but also because the initial distribution that minimizes $\Delta \mathcal{S}^\text{irr}$ is similar to the one that minimizes $D_{KL}(P^f_M || P^i_M)$. However, the minimal copy time $\tau_{\rm min}$ for copying the unlikely state is much larger for $p_1(0) = \avg{p_1(\tau)}$ than for $p_1(0)=1/2$ (see Fig. S4). Concomitantly,  for short copy times $\tau$, the cost $W_d$ to copy the unlikely state of the data bit, and thereby  $\Delta \mathcal{S}^\text{irr}$, rises dramatically if $p_1(0)$ remains equal to $\avg{p_1(\tau)}$, (\fref{main}(a)). 
Indeed, in this regime the optimal value of $p_1(0)$ moves toward $p_1(0)=1/2$ (Fig. \ref{fig:main}(a,g)). The analytical $1/\tau$ approximation of $\cal{W}$ \cite{SI} does not predict this shift of $p_1^{\rm opt}(0)$ (dashed lines \fref{main}(a)).

\begin{figure}
    \centering
    \includegraphics[width=\columnwidth]{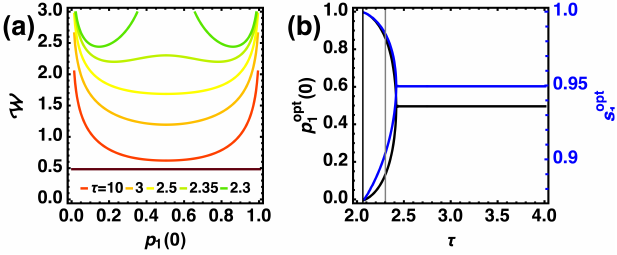}
    \caption{Unequal copying accuracies induce symmetry breaking at high copy speeds for unbiased data bit ($p'=0.5$). (a) $\mathcal{W}$ as function of $p_1(0)$ for different copy times $\tau$; $s_1$ has been optimized. (b) Optimal $p_1(0)$ and $s_1$ as function of $\tau$. Black vertical line: $\tau_{\rm min}$ for $s_0\neq s_1$; grey vertical line for $s_0=s_1$. }
    \label{fig:sb}
\end{figure}

Fig. \ref{fig:main}(b)-(f) shows how this picture is fundamentally
transformed if different relative copy accuracies $s_0,s_1$ can be
used to reduce the average work $\mathcal{W}$ at fixed $I(D;M)$. These
plots report $\cal{W}$ in the quasi-static limit (panel (c)) and at finite-times
 ((d)-(f)). As $\tau$ decreases, the
difficulty of moving probability between memory states renders large
regions of $p_1(0)-s_1$ space inaccessible. For $s_0=s_1$ (grey
vertical line), the optimal value of $p_1(0)$ is thus forced towards
$1/2$. However, by increasing $s_1$ and decreasing $s_0$ along the
iso-information contour (Fig. \ref{fig:measurementfinitetime}(b)), a
wedge of low cost copying opens up, because the cost of copying the
unlikely state $d=0$ decreases more than that of $d=1$ rises, for two
reasons:
$ (1-p') |\partial W_0 / \partial s_0| > p' |\partial W_1 / \partial
s_1|$, and along the iso-information line $s_0$ falls more than $s_1$
rises, $|ds_0|>|ds_1|$.  Now $p_1^\text{opt}(0)$ moves away from
$1/2$, ending above $p_1(0)=\avg{p_1(\tau)}$ (red line), see also
Fig. \ref{fig:main}(g). As a result, not only
 $\mathcal{W}$ is lowered (\fref{main}(b)), but also the minimum copy
 time $\tau_{\rm min}$ (\fref{main}(g)). Allowing for unequal copying accuracies thus lowers the average work, especially when $\tau$ is small, and raises the maximum copy speed (see also Fig. S5).

We have seen that optimal copying at fixed $I(D;M)$ tends to induce a  bias in the initial memory distribution that is stronger than that present in the data. \fref{sb} shows that the same effect can even induce a symmetry breaking transition for an unbiased data bit ($p'=0.5$): when $\tau$ is long, the optimal initial memory distribution is symmetric and both data bit states are copied with the same accuracy, yet for sufficiently short copy times, this distribution is pushed to either 0 or 1, with one data bit state being copied more accurately than the other. The symmetry breaking drastically lowers the work and raises the maximum copy speed (\fref{sb} and Figs. S8 and S9).

The procedure considered here leads to irreversible entropy production not only because of the finite-time cost $T \Delta \mathcal{S}^{\rm irr}$, but also because it neither recovers the free energy stored in the correlations between data and memory bit, $k_B T I(D;M)$,  nor that stored in the marginal distribution of the memory after the copy process, $k_BT (D_{KL}(P_M^f||P_M^i)$ (\eref{Wtot}). While recovery of $k_B T I(D;M)$ requires knowledge of the state of the data bit, retrieval of  $k_BT (D_{KL}(P_M^f||P_M^i)$ only requires $p'$. Recovering the latter, i.e. the free energy stored in the marginal memory distribution, underlies procedure II, with  total cost  $\mathcal{W}_{\rm we}= k_BT I(D;M) + T \Delta
  \mathcal{S}^\text{irr}$. While this
  extraction lowers the cost, with $p_1^{\rm opt}(0)$ pushed even further beyond $\avg{p_1(\tau)}$ (Fig. S7), it requires a time-consuming
  decorrelation step. If the full cycle is under a time constraint,
  then it is optimal to directly overwrite the memory \cite{SI}.  When
    successive data bits are uncorrelated, such that $I(D;M)=0$ at the
    start of the cycle, the overall cost is
  $\mathcal{W}_{\rm do}= k_BT I(D;M) + T \Delta
  \mathcal{S}^\text{irr}$; the constraint $p_1(0)=\avg{p_1(\tau)}$
  makes the cost and minimum copy time $\tau_{\rm min}$ higher than
  those of the other procedures (Fig. S7).

\textit{Conclusion -} A known bias in a data bit can be leveraged to reduce the cost of copying it. Generally, it is advantageous to match the initial memory distribution to the expected copy outcome. However, when operating close to the memory's relaxation time, this rule of thumb trades off against the cost of moving a probability distribution quickly. Optimal protocols are therefore pushed away from the rule of thumb, but in a way that strongly depends on how the protocol is constrained: for short copy times, the optimal protocol for copying at equal accuracies is different from that at unequal accuracies. The latter  dramatically lowers the work, raises the maximum copy speed, and can even induce an unexpected symmetry breaking transition. These observations may be used to optimize copy operations in Random Access Memories and multi-state memories \cite{Remlein.2021}, and enhance  more complex computing systems like logic gates \cite{Wolpert2020}. Our results raise the possibility that algorithms that perform more likely operations with higher accuracy may be beneficial for building low cost computers.

\textit{Acknowledgements -} We thank Avishek
Das for a careful reading of the manuscript.
This work is part of the Dutch Research Council (NWO) and was performed
at the research institute AMOLF. This project has received funding from the
European Research Council under the European Union’s Horizon 2020 research
and innovation program (grant agreement No. 885065). TEO supported by a Royal Society University Research Fellowship.

\end{document}